\documentclass[review]{elsarticle}

\usepackage{lineno,hyperref}
\usepackage[margin=25mm]{geometry}
\usepackage{siunitx}
\usepackage{amsmath,amssymb}
\usepackage{booktabs}
\usepackage{nomencl}
\modulolinenumbers[2]
\makenomenclature
\usepackage{etoolbox}
\renewcommand\nomgroup[1]{%
  \item[\bfseries
  \ifstrequal{#1}{A}{Roman characters}{%
  \ifstrequal{#1}{B}{Greek characters}{%
  \ifstrequal{#1}{C}{Subscripts}{}}}%
]}

\newcommand{\nRe}{\mathord{\mathit{Re}}}
\newcommand{\nPr}{\mathord{\mathit{Pr}}}
\newcommand{\degC}{\si{\degreeCelsius}}
\newcommand{\sidegC}[1]{\SI{#1}{\degreeCelsius}}

\journal{International Communications in Heat and Mass Transfer}

\usepackage{numcompress}

\begin{document}

\begin{frontmatter}

\title{Temperature depression model for cavitating flow with thermodynamic suppression effect in high-temperature water}

\author[IFS]{Junnosuke Okajima\corref{mycorrespondingauthor}}
\cortext[mycorrespondingauthor]{Corresponding author}
\ead{j.okajima@tohoku.ac.jp}
\author[IFS,ENG]{Taku Hanyuda}
\author[IFS]{Yuka Iga}

\address[IFS]{Institute of Fluid Science, Tohoku University, 2-1-1 Katahira, Aoba-ku, Sendai, Miyagi 980-8577, Japan}
\address[ENG]{Mechanical Engineering Division, School of Engineering, Tohoku University, 6-6, Aramaki Aza Aoba, Aoba-ku, Sendai, Miyagi 980-8579, Japan}

\begin{abstract}
The thermodynamic suppression effect of cavitation generally appears in cryogenic cavitating flows. 
Temperature depression, which is the temperature difference between the mainstream and inside the cavity, indicates the thermodynamic suppression effect.
In this study, a temperature depression model is developed to understand the physical process of the thermodynamic suppression effect. The model is evaluated using the experimental data of temperature inside supercavitation in high-temperature water of up to \sidegC{140}.
The temperature depression model is derived based on Fruman's model and newly introducing the suppression effect of evaporative mass flux. 
At first, Fruman's  model was compared with the experimental data.
Fruman's model differed from the experimental data in the high-temperature region of more than 100 \degC. In contrast, the proposed model reproduced the experimental data well in the high-temperature region.
Therefore, introducing a suppression effect of evaporative mass flux is essential for describing the temperature depression in cavitating flow.
In addition, the proposed model was expressed with existing parameters, and it was clear that the temperature depression was defined with the characteristic temperature of the B factor, the Nusselt number, and a term representing the suppression effect of evaporative mass flux expressed by dimensionless thermodynamic parameters. 
\end{abstract}

\begin{keyword}
multiphase flow \sep cavitation \sep phase change \sep NACA0015 \sep thermodynamic suppression effect
\end{keyword}

\end{frontmatter}


\section{Introduction}
The thermodynamic suppression effect of cavitation generally appears in cavitating flows in cryogenic fluids \cite{Stahl1956, Stepanoff1964}, fluorocarbons\cite{Franc2004, Fruman1999}, and hot water\cite{Kato1996}; this effect is preferable for fluid machinery because inception and development of cavitation should be suppressed.
This effect can be essential in improving the performance of a rocket engine using cryogenic fluid as a propellant and realizing a liquid hydrogen pump for the hydrogen economy.
Therefore, extensive experimental studies and numerical modeling\cite{Le2019} of this effect in cavitating flows have been conducted. 

Cavitation generally responds to thermal stimulation, as demonstrated in the heated hydrofoil experiment \cite{Okajima2022, Yang2023}; therefore, the thermodynamic suppression effect is mainly caused by the thermal characteristics determined by the thermophysical properties.
The mechanism of the thermodynamic suppression effect can be explained as follows; 1. The sensible heat in the liquid around the vapor bubble is consumed by evaporation during cavity development, 2. The temperature of the liquid-vapor interface decreases, 3. The saturation pressure also decreases, and evaporation is suppressed. 
Thus, this effect affects cavitation itself, so it is an intrinsic characteristic of the working fluid.
The thermodynamic parameter proposed by Brennen\cite{Brennen1973} indicates this effect and is expressed as follows:

\begin{equation}
	\Sigma = \frac{\rho_V^2 L^2}{\rho_L^2 c_{pL} T_\infty \sqrt{\alpha_L}}.
	\label{eq_dim_Sigma}
\end{equation}

\noindent
This parameter generally increases rapidly as the liquid temperature approaches its critical temperature, and the effect becomes noticeable.

Temperature depression, which is the temperature difference between the mainstream and inside the cavity, indicates a thermodynamic suppression effect.
A more significant temperature depression indicates that the cavity size is significantly suppressed by evaporation.
Hence, studies have been conducted on temperature measurements inside cavities.
Hord\cite{Hord1973} measured the cavity temperature in liquid hydrogen and nitrogen using thermocouples installed on a hydrofoil surface.
These measurements are the only available data for cavitating flow in liquid hydrogen.
Niiyama et al.\cite{Niiyama2012} measured the cavity temperature of liquid nitrogen using a diode sensor and discussed the variation in the thermal boundary layer thickness around the cavitation bubble for Reynolds number.
Petkovs\v{e}k and Dular\cite{Petkovsek2017} measured the temperature field in cavitating flow using an infrared camera.
Our group developed a measurement system for temperature depression using a thermistor probe located in the fluid. Furthermore, we investigated the thermodynamic suppression effect of tip-vortex cavitation\cite{Kang2019} and cavitation on NACA0015 in water at 80 \degC\cite{Iga2016,OKUBO2022}.

The first model to describe the amount of temperature depression was first proposed by Holl et al.\cite{Holl1975} followed by Kato\cite{Kato1984} and Fruman et al.\cite{Fruman1999}.
Our study indicates that Fruman's model could describe the temperature depression of supercavitation in hot water of up to 80 \degC\cite{Iga2016,Iga2022}.
We also demonstrate that the temperature depression estimated from the cavity length variation agrees with Fruman's model\cite{Iga2022}.
However, the model has yet to be validated based on temperature measurements in the cavitating flow of more than 100 \degC, where the thermodynamic suppression effect is strong.
Hot water of \SI{100}{\degreeCelsius} and above has a suppression effect similar to liquid nitrogen. Hence, it is helpful to investigate the thermodynamic suppression effect. However, the available data are limited after the last experiment by Kato et al.\cite{Kato1996}.
Therefore, we recently conducted experiments using hot water at temperatures greater than 100 \degC\cite{HANYUDA2021}.
In this study, we propose a new model describing temperature depression in cavitating flow and validate the model using temperature measurement data of supercavitation around NACA0015 of up to 140 \degC.

\section{Modeling of temperature depression}
\begin{figure} [htbp]
\centering
	\includegraphics[scale=0.8]{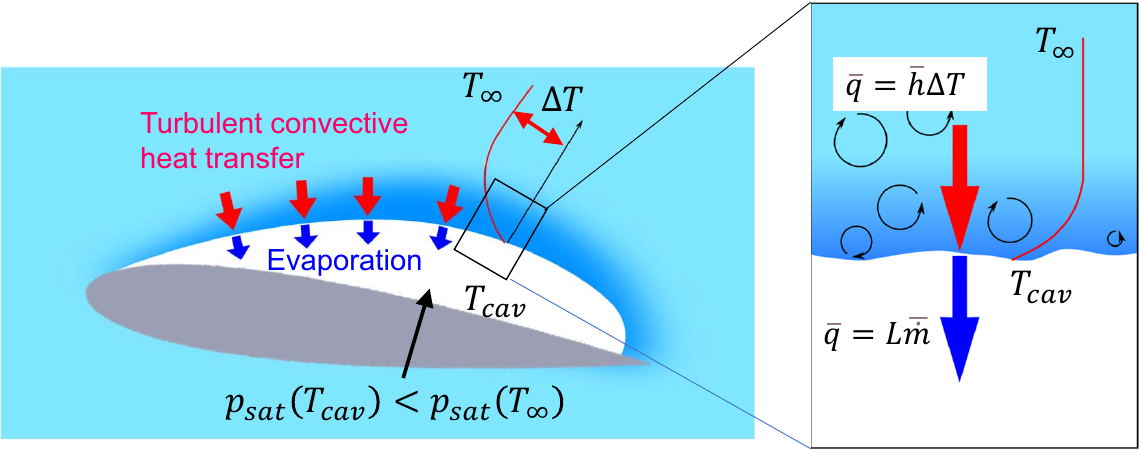}
\caption{Schematic of heat transfer process on the cavitation surface}
\label{fig_model}
\end{figure}

The basic concept for describing the temperature depression is shown in Fig. \ref{fig_model}. The energy balance at the liquid-vapor interface is expressed as

\begin{align}
	\overline{q_{evp}} &= \bar{h} \left( T_\infty - T_{cav} \right) \notag \\
	              &= \bar{h} \varDelta T.
	\label{eq_energy_balance}
\end{align}

\noindent
To describe the actual phenomena, the modeling of the average heat transfer coefficient $\overline{h}$ and evaporative heat flux $\overline{q_{evp}}$ are required.

\subsection{Model by Fruman et al.~\cite{Fruman1999}}
Fruman et al. described temperature depression based on the relationship between convective heat transfer and evaporative heat flux at the cavity interface.
They assumed that 1. The cavity surface is a rough flat plate, 2. The temperature inside the cavity is uniform, and 3. The evaporative mass flux is proportional to the mainstream velocity.
The cavity temperature, which was considered the wall temperature in the model, was expressed using Reynolds analogy, and the evaporative heat flux was defined as

\begin{equation}
	 \overline{h} = \frac{\displaystyle\frac{1}{2} C_f \rho c_p U_\infty}{1+\displaystyle \frac{U_b}{U_\infty} \left(\nPr-1\right)},
	\label{eq_Tw_Re_analogy}
\end{equation}
\begin{equation}
	\overline{q_{evp}} = L \overline{\dot{m}_0} = \rho_v L C_Q U_\infty,
	\label{eq_q_mdot_0}
\end{equation}

\noindent
where the evaporative mass flux $\overline{\dot{m}_0}$ is assumed to be proportional to the mainstream velocity.
By applying the turbulent boundary layer relationship, the following equation is obtained:

\begin{equation}
	\varDelta T = \frac{\displaystyle \rho_V L C_Q}{\displaystyle 0.00695 \rho_L c_{pL}}
	\left( \frac{\displaystyle x}{\displaystyle \epsilon} \right)^{\frac{1}{7}}
	\{ 1-2.1 \nRe_x^{-0.1} \left( 1- \nPr \right) \}
	\label{eq_Fruman}
\end{equation}

\subsection{Proposed model}
In this study, the energy balance on the cavity surface was considered similar to Fruman's approach. 
The main difference is that our model includes the thermodynamic suppression effect in the derivation of evaporative mass flux.
The local evaporative mass flux is assumed to be proportional to the difference between the local pressure $p$ and local saturation pressure $p_{sat}$, and is expressed as

%

\begin{equation}
	\dot{m} = C_e \left( p_{sat} \left(T_{cav} \right) - p \right)
	\label{eq_mdot_base}
\end{equation}

\noindent
where $C_e$ is a coefficient related to the evaporation rate, the temperature inside the cavity determines the saturation pressure. 
Hence, evaporative heat flux also depends on the mainstream and cavity temperatures, as expressed by

\begin{equation}
	\overline{q_{evp}} = L \overline{\dot{m}},
	\label{eq_q_mdot}
\end{equation}
Equations \ref{eq_energy_balance}, \ref{eq_mdot_base}, and \ref{eq_q_mdot} are the fundamental equations of the proposed temperature–depression model.

When the temperature depression $\varDelta T$ is sufficiently small, the expansion of the saturation pressure in Eq. \ref{eq_mdot_base} around the mainstream temperature $T_\infty$ is expressed as,

\begin{align}
	\dot{m} &= C_e \left( p_{sat} \left(T_{\infty} \right)- \frac{d p_{sat}}{dT}\varDelta T - p \right) \notag \\
	        &= C_e \left( p_{sat} \left(T_{\infty} \right) - p \right) - C_e \frac{d p_{sat}}{dT}\varDelta T,
	\label{eq_mdot_transform}
\end{align}

\noindent
Integrating Eq.\ref{eq_mdot_transform} on the cavity surface and dividing the surface area of the cavity gives the average evaporative mass flux.
The first term on the right-hand side of Eq.\ref{eq_mdot_transform} represents the evaporation rate in isothermal cavitating flow and is determined by the flow field.
The average evaporative mass flux in isothermal cavitating flow is denoted by $\overline{\dot{m}}$, which is the same parameter as in Eq.\ref{eq_q_mdot_0}.
Assuming that the temperature inside the cavity is uniform, the second term on the right-hand side of Eq.\ref{eq_mdot_transform} is not changed by integration and averaging treatment.
Subsequently, the average evaporative mass flux is expressed as follows:

\begin{equation}
	\overline{\dot{m}}  = \overline{\dot{m_0}} - C_e \frac{d p_{sat}}{dT} \varDelta T
	\label{eq_mdotbar}
\end{equation}

Eq.\ref{eq_mdotbar} can be substituted into Eq.\ref{eq_q_mdot}.
In addition, by introducing Fruman's assumption of evaporative mass flux in isothermal cavitating flow shown in Eq.\ref{eq_q_mdot_0} and introducing the Clapylon-Clausius equation into the second term, the evaporative mass flux can be expressed as,

\begin{equation}
	\overline{q_{evp}} = C_Q \rho_V L U_\infty - C_e \frac{\rho_V L^2}{T_\infty} \varDelta T.
	\label{eq_qbar_suppression}
\end{equation}

Subsequently, by introducing Colburn's analogy and assuming that the friction factor on the cavity surface is proportional to $Re^{- \frac{1}{5}}$, Eq.\ref{eq_energy_balance} can be expressed as,

\begin{equation}
	\overline{q_{evp}} \sim \frac{k}{L_c} Re^{\frac{4}{5}} Pr^{\frac{1}{3}} \varDelta T.
	\label{eq_qbar_Colburn}
\end{equation}

By combining Eqs.\ref{eq_qbar_suppression} and \ref{eq_qbar_Colburn}, $\varDelta T$ can be rewritten as,

\begin{equation}
	\varDelta T = \frac{\displaystyle \rho_V L}{\displaystyle C_t \rho_L c_{pL} Re^{-\frac{1}{5}} Pr^{-\frac{2}{3}}+ C_s \frac{\rho_V L^2}{U_\infty T_\infty}},
	\label{eq_dT_dimension}
\end{equation}

\noindent
where $C_t$ and $C_s$ are the convective heat transfer and evaporation suppression coefficients, respectively.
Compared to Fruman's model in Eq.\ref{eq_Fruman}, one term is added to the denominator.
The right-hand sides of Eq.\ref{eq_dT_dimension}, the numerator, the first and second terms in the denominator represent the degree of temperature decrement by evaporation, the degree of turbulent convective heat transfer at the cavity surface and the reduction degree of evaporative mass flux due to temperature decrement in the cavity, respectively.
Hence, the second term in the denominator reflects the influence of the suppressed amount of cavity by the thermodynamic suppression effect.
Because the first and second terms of the denominator are positive, when these terms are larger, $\Delta T$, namely the thermodynamic suppression effect, becomes weaker. In contrast, the effect becomes stronger when the numerator $\rho_V L$ is larger.
Therefore, equation \ref{eq_dT_dimension} indicates that a large turbulent convective heat transfer reduces the temperature difference between the mainstream and cavity, and a large suppression effect reduces the temperature depression because of the reduction in evaporative mass flux.

\section{Experimental system}
\begin{figure} [htbp]
\centering
	\includegraphics[scale=1.2]{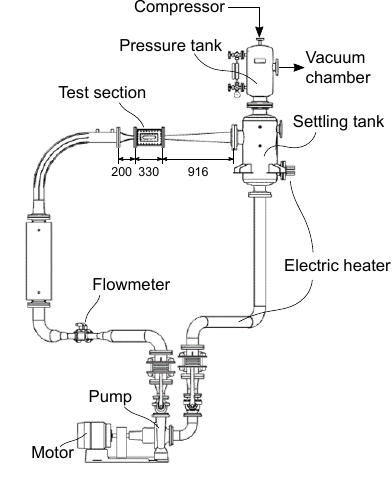}
\caption{Schematic of high-temperature water cavitation tunnel}
\label{fig_tunnel}
\end{figure}

The experiments were conducted in a high-temperature water cavitation tunnel at the Institute of Fluid Science, Tohoku University, Japan.
This experimental apparatus was also used in our previous studies \cite{Iga2016, Okajima2022, Iga2022}.
Figure \ref{fig_tunnel} shows a schematic diagram of the abovementioned apparatus.
The test section was a rectangular channel with a span, height, and length of 20, 30, and 330 mm, respectively.
The basic specifications incorporated a flow rate of 700 L/min, maximum operation pressure of 0.51 MPa, and maximum operating temperature of \SI{140}{\degreeCelsius}.
The settling tank was connected to a compressor and vacuum pump, and the operating pressure was controlled from negative to pressurized conditions.
The mainstream pressure was measured using a pressure transducer (PH-10KB, Kyowa Electric Instruments Co., Ltd.) located 110 mm upstream from the center of the hydrofoil.
The mainstream temperature was controlled using two electric heaters with an accuracy of $\pm$ 0.1 K.
Therefore, several cavitation patterns can be evaluated from room temperature to 140 \degC.

\begin{figure}[htbp] 
\centering
	\includegraphics[scale=0.5]{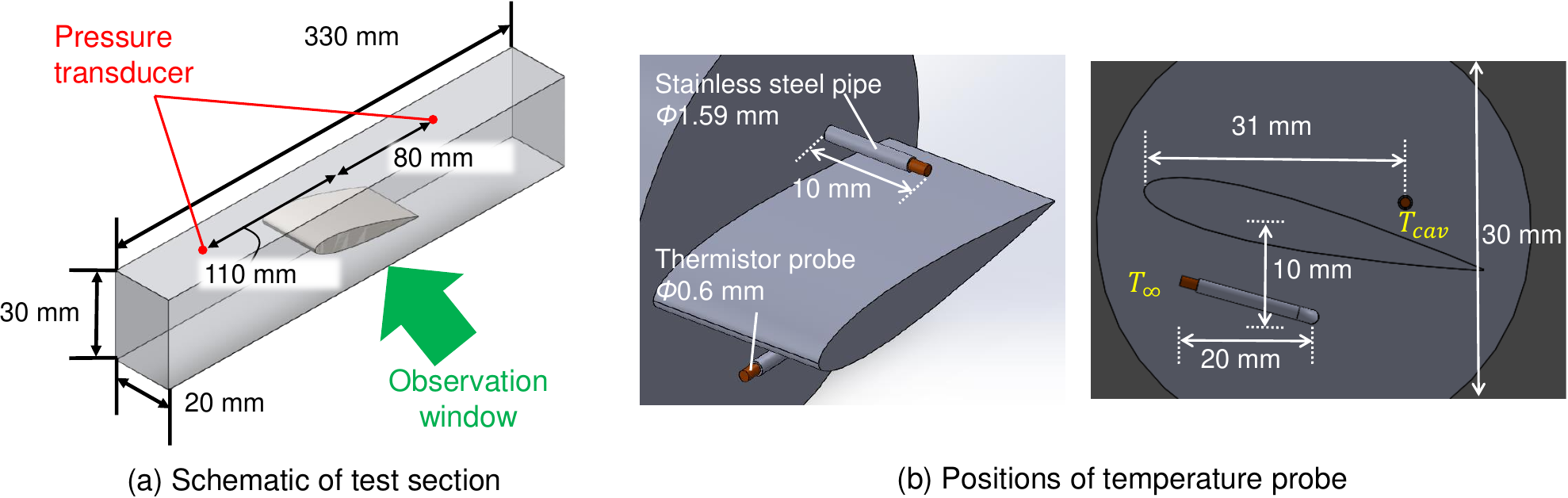}
\caption{Schematic of test section }
\label{fig_testsect}
\end{figure}

Figure \ref{fig_testsect} illustrates the details of the test section.
NACA0015 hydrofoil with a chord length of 40 mm and a span of 20 mm was used.
The angle of attack and blockage ratio were 12 deg. and 0.33, respectively.
In Fig. \ref{fig_testsect}, the locations of the temperature sensors are also shown.
The cavity temperature $T_{cav}$ and mainstream temperature $T_\infty$ were measured using the temperature sensor located on the suction and pressure sides of the hydrofoil, respectively.
A thermistor probe (N317/BR14KA103K/23300/RPS/3/SP, Nikkiso-Thermo Co., Ltd) was used as the temperature sensor.
The thermistor chip was molded with epoxy resin in a polyimide tube with a diameter of 0.6 mm.
The time constant of this probe was 200 ms for a 63 \% response in a liquid.
To hold the temperature sensor in the cavitating flow, the thermistor probe was inserted into a stainless-steel tube of diameter 1.59 mm and fixed to the test section.
The electrical resistance of the thermistor was measured using a digital multimeter (DMM4040, Tektronix Co., Ltd). 

Each thermistor was calibrated using a standard thermometer consisting of a secondary standard platinum-resistant thermometer (5616-12, Fluke Co., Ltd.) and a Black Stack thermometer readout (1560, Fluke Co., Ltd.).
The temperature sensors were put inside the aluminum block submerged in oil inside a Dewar vessel to ensure a uniform temperature distribution around the sensors.
After heating the system inside the Dewar vessel using a rubber heater, the temperature of the system gradually decreased by releasing the heat into the ambient environment.
Finally, the expanded uncertainty was 16 mK at 40 \degC, 19 mK at 80 \degC, and 29 mK at 140 \degC.

\begin{figure}[htbp] 
\centering
	\includegraphics[scale=0.7]{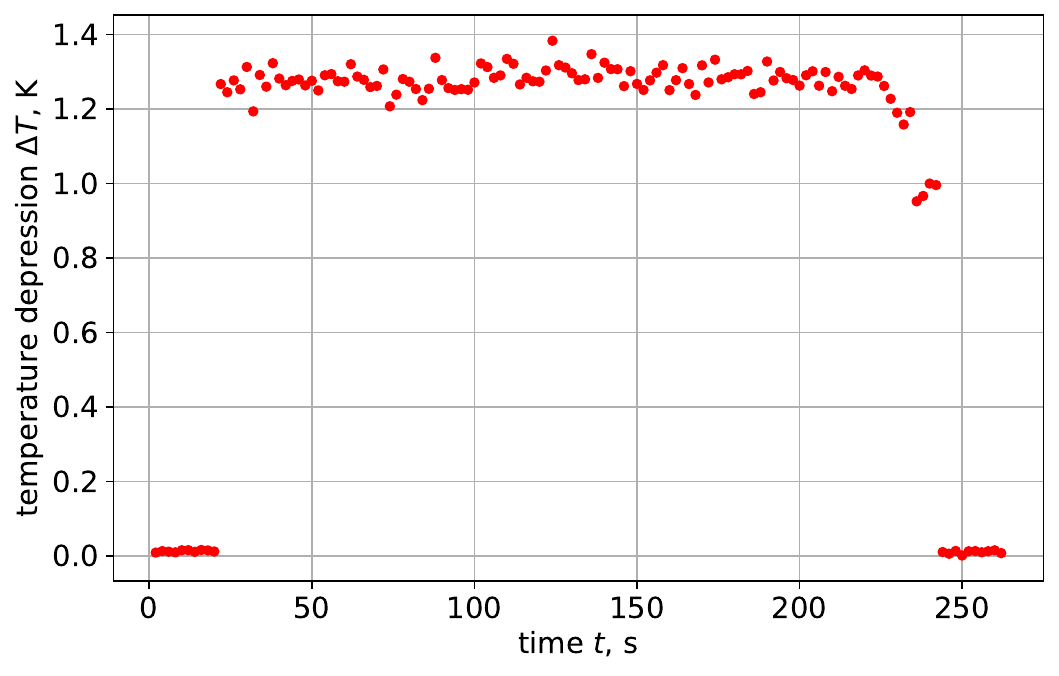}
	\caption{Time variation of temperature depression in 140 \degC}
\label{fig_t_dT}
\end{figure}

Fig. \ref{fig_t_dT} shows an example of temperature measurement data.
In this case, the mainstream temperature was 140 \degC.
After logging began, supercavitation was generated by adjusting the mainstream pressure, and the cavity reached the temperature sensor within 20 s.
The temperature inside the supercavitation was almost stable, with a fluctuation of approximately 0.2 K.

\begin{table}[htbp]
\caption{Experimental conditions}
\begin{center}
\label{tab_exp_cond}
\begin{tabular}{ll} \toprule
	Mainstream temperature $T_\infty$, \si{\degreeCelsius} & 50--140 \\
	Mainstream velocity $U_\infty$, m/s & 13.6 \\
	Angle of attack $\alpha_{in}$, \si{\degree} & 12 \\
	Cavitation number $\sigma$, - & 1.0 \\ \bottomrule
\end{tabular}
\end{center}
\end{table}

\section{Results and discussion}
\begin{figure}[htbp] 
\centering
	\includegraphics[scale=0.7]{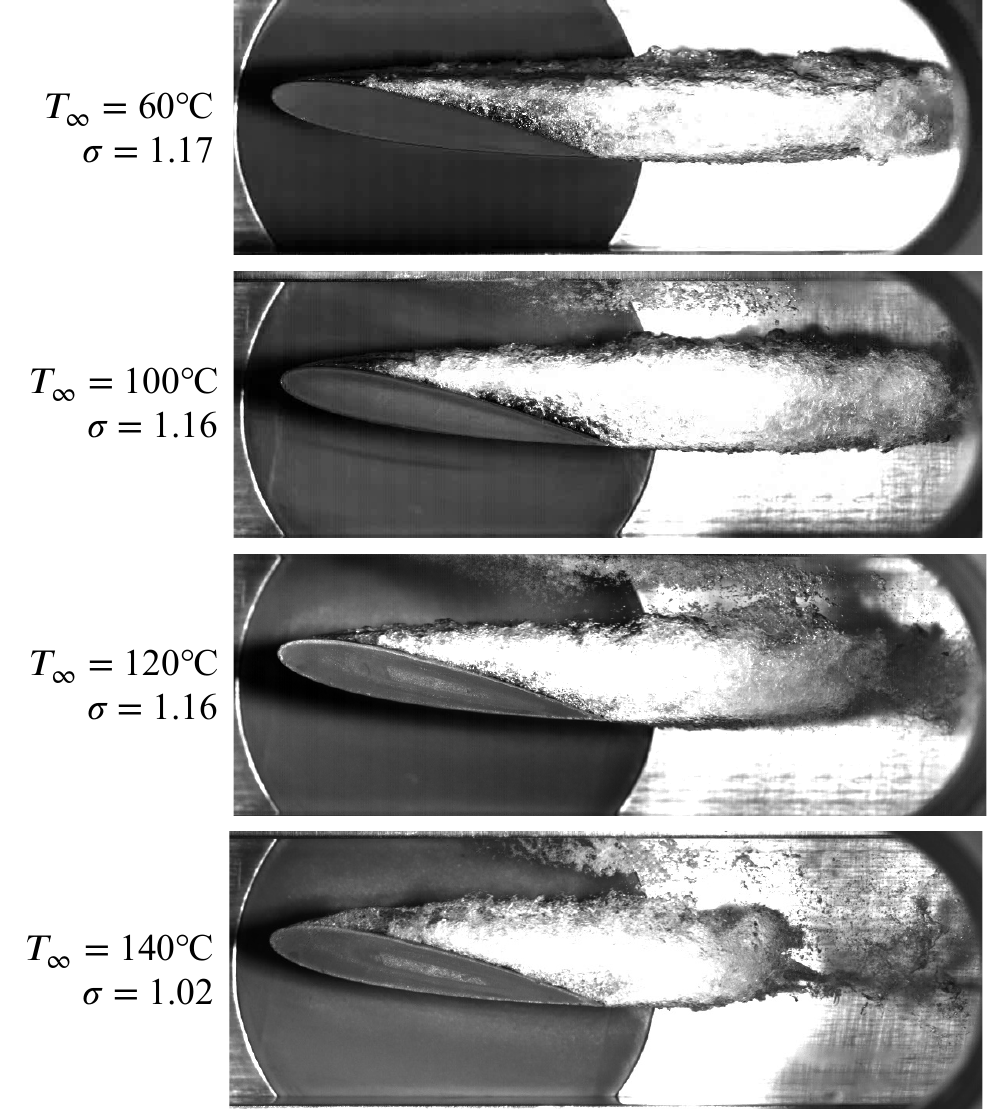}
\caption{Aspects of supercavitation with various mainstream temperature}
\label{fig_aspct}
\end{figure}

The cavity aspects are shown in Fig. \ref{fig_aspct}.
In these images, each cavitation number differed from the temperature measurement because the cavities were longer than the observation window.
As shown in Fig. \ref{fig_aspct}, the cavity became smaller as the mainstream temperature increased.
Significantly, at a mainstream temperature of \ SI{140}{\ degreeCelsius}, the cavity was more strongly suppressed even lower mainstream temperature than in the other cases.
These images imply that the evaporative mass flux decreases as the mainstream temperature increases.

\begin{figure}[htbp] 
	\centering
	\includegraphics[scale=0.7]{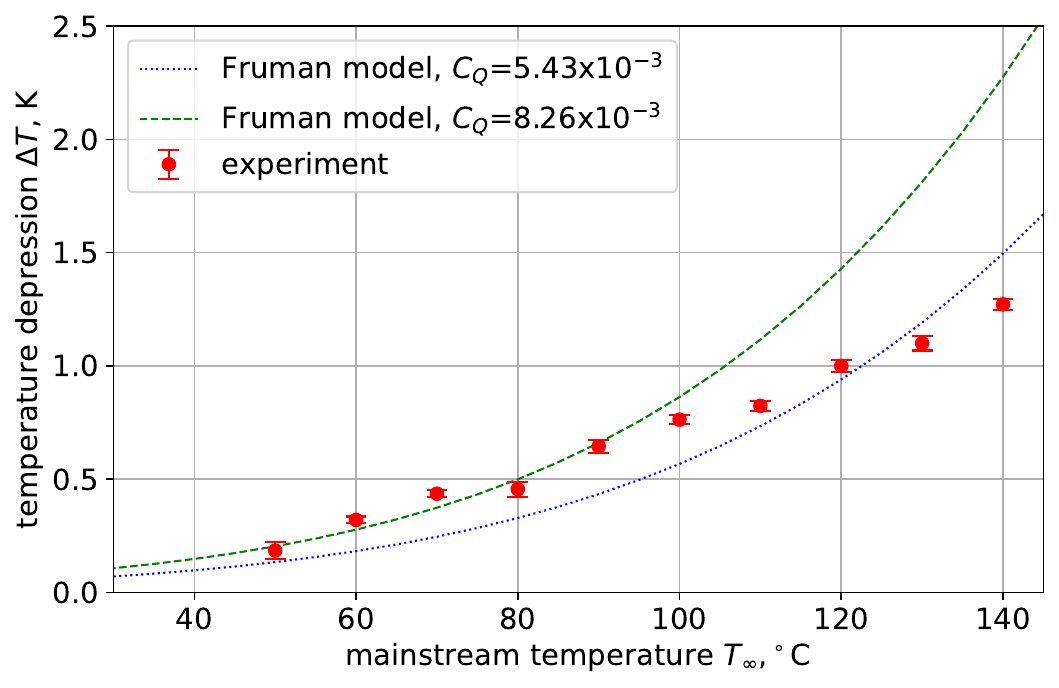}
	\caption{Relationship between temperature depression inside the cavity and mainstream temperature at constant $\sigma$ ($\sigma = 1.0$) and comparison with temperature depression model by Fruman et al.\cite{Fruman1999}}
	\label{fig_dT_Fruman}
\end{figure}

Figure \ref{fig_dT_Fruman} shows the variation in the temperature depression $\Delta T$ inside the cavity against $T_\infty$ at a constant $\sigma$ in which the experimental results in this study and the predicted value from Fruman's model of Eq. \ref{eq_Fruman} \cite{Fruman1999} are compared.
The experimental coefficient $C_Q$ was determined to be $C_Q = 5.43 \times 10^{-3}$ when Fruman's model was fitted to the entire set of experimental data. 
As shown in Fig. \ref{fig_dT_Fruman}, the model with this coefficient does not agree with the experimental data trend.
In addition, similar to a previous study \cite{Iga2022}, curve fitting was performed on the experimental data up to 80 \degC, and $C_Q = 8.26 \times 10^{-3}$ was obtained.
In this mainstream temperature range, Fruman's temperature-depression model reproduced the experimental data well; however, the difference became more significant at higher temperatures.
Fruman's model assumes that the evaporation rate is only proportional to the mainstream velocity and does not consider the decrement in evaporation rate at a high mainstream temperature.

\begin{figure}[htbp] 
\centering
	\includegraphics[scale=0.7]{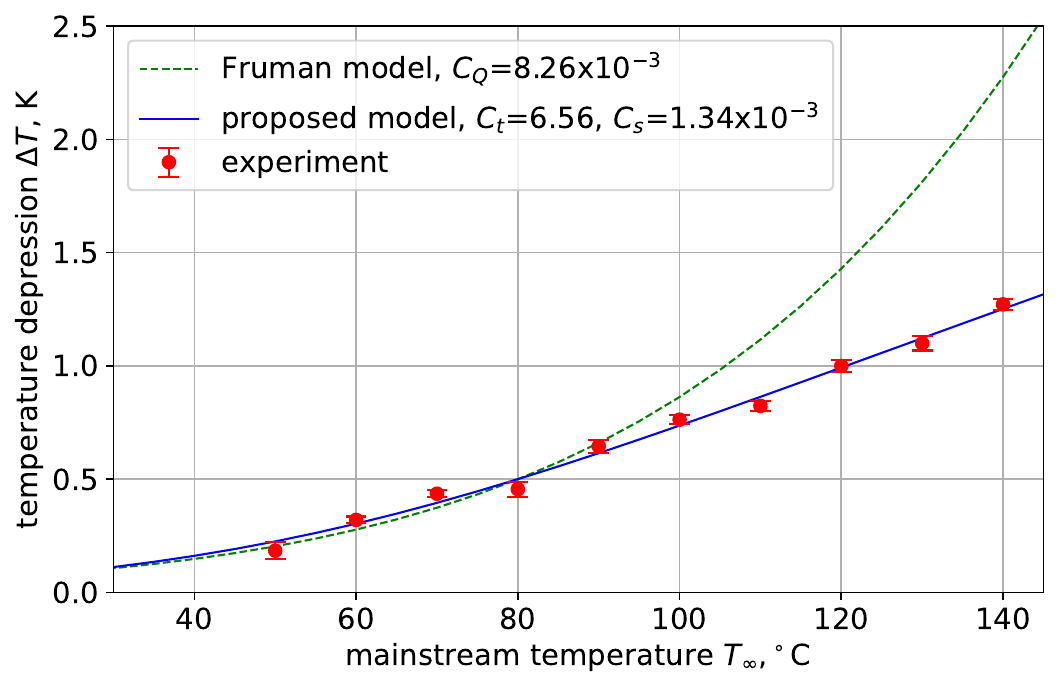}
	\caption{Comparison of temperature depression inside the cavity and mainstream temperature at constant $\sigma$ ($\sigma = 1.0$) among experimental data, temperature depression model by Fruman et al. with $C_Q$ = 8.26×10-3, and proposed temperature depression model with $C_t$ = 6.56 and $C_s$ = 1.34×10$^{-3}$}
\label{fig_dT_new}
\end{figure}

Figure \ref{fig_dT_new} compares the proposed model and Fruman's temperature–depression model.
The coefficients $C_t$ and $C_s$ in Eq. \ref{eq_dT_dimension} were determined using the modified Powell method to minimize the difference from the experimental data.
In contrast to the values predicted by Fruman's model up to \SI{80}{\degreeCelsius} of the mainstream temperature overestimated in the high-temperature range, the proposed model accurately reproduced the temperature variation at high temperatures.
Therefore, introducing the suppression effect of evaporative mass flux is essential for describing the temperature depression in cavitating flow.

Iga et al.\cite{Iga2022} measured the cavity length arising on NACA0015 in high-temperature water and estimated the temperature depression using the quantity of thermodynamic suppression effect $|\Delta \sigma|_{l_{cav}}$ evaluated from the variation of the cavity length.
The temperature depression expressed with $|\Delta \sigma|_{l_{cav}}$ as;
\begin{equation}
	\varDelta T = \frac{U^2_{\infty} T}{2 L} \frac{\rho_L}{\rho_V} |\Delta \sigma|_{l_{cav}}
	\label{eq_dT_sigma_Lcav}
\end{equation}
In their study, the temperature depression at a mainstream temperature higher than \SI{100}{\degreeCelsius} was estimated by extrapolation using Eq. \ref{eq_dT_sigma_Lcav}, and Fruman's model reproduced the estimated temperature depression trend.
This study showed that the actual temperature depression in the cavitating flow around the NACA0015 hydrofoil, which has the same experimental configuration as Iga et al.\cite{Iga2022}, was smaller than the values predicted by Iga et al.\cite{Iga2022} using Fruman's model.
The estimation method using Eq. \ref{eq_dT_sigma_Lcav} assumes that the variation of saturation pressure causes the cavity length variation because of temperature depression and does not consider the reduction in evaporation rate.
Therefore, the estimated temperature depression should be overestimated compared to the experimental data.

\begin{figure}[htbp] 
\centering
	\includegraphics[scale=0.7]{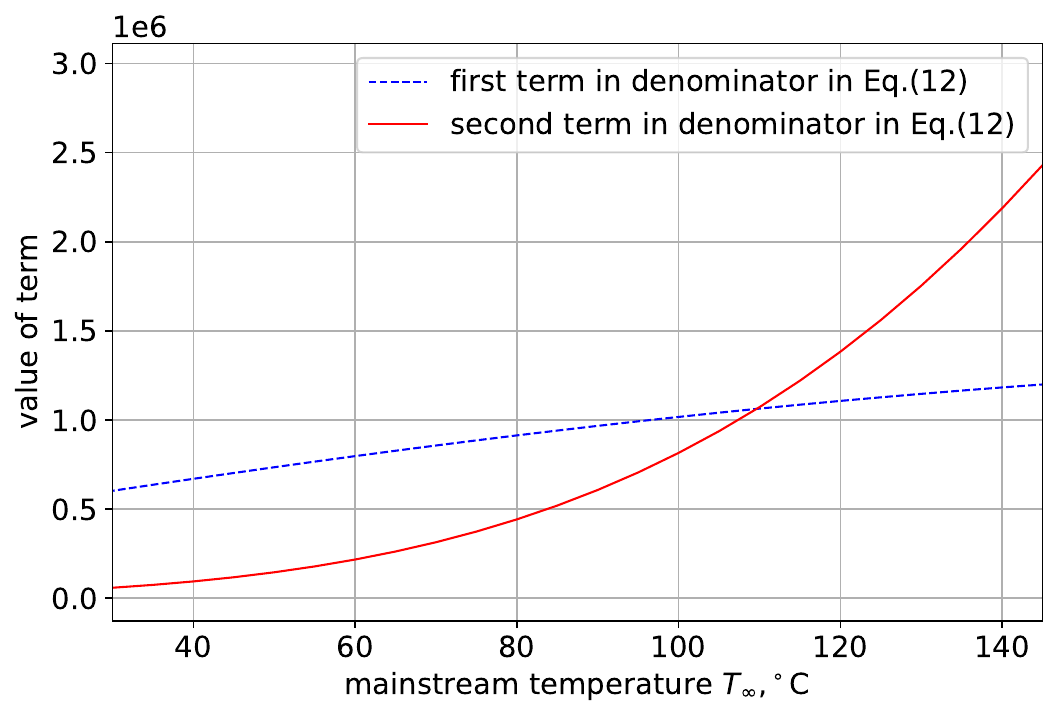}
	\caption{Variation of first and second terms of the denominator in the proposed temperature depression model Eq. (\ref{eq_dT_dimension}) in the case of fitting to the present experimental result}
\label{fig_compose}
\end{figure}

Figure \ref{fig_compose} compares the terms in the denominator of Eq. \ref{eq_dT_dimension}.
The two lines intersected at the mainstream temperature of 110 \degC.
When the mainstream temperature is lower than the intersection, the first term is larger than the second term; that is, turbulent convective heat transfer is dominant.
Fruman's model considers only the first term in the denominator.
The first term is weakly dependent on the mainstream temperature.
The numerator $\rho_V L$ increases much more strongly than the first term of the denominator as the mainstream temperature increases; therefore, Fruman's model overestimates the temperature depression in the high-temperature region.
The second term of the denominator rapidly increased in high-temperature regions.
This term suppresses the increase in the temperature depression governed by the temperature dependency of $\rho_V L$.
The intersection of the two lines is a candidate indicator for the onset of thermodynamic suppression.

The derived temperature depression Eq. \ref{eq_dT_dimension} is transformed by known parameters.
Stepanoff's B factor \cite{Stepanoff1964} describes the strength of the thermodynamic suppression effect and is expressed as follows:

\begin{equation}
	B = \frac{\varDelta T}{T^*}, T^* = \frac{\rho_V L}{\rho_L c_{pL}}.
	\label{B_factor}
\end{equation}

\noindent
The dimensionless thermodynamic parameter introduced by Brennen\cite{Brennen1995} is expressed as follows:

\begin{equation}
	\Sigma^* = \frac{\rho_V^2 L^2}{\rho_L^2 c_{pL} T_\infty \sqrt{\alpha_L}} \sqrt{\frac{L_c}{U_\infty^3}}.
	\label{dimless_Sigma}
\end{equation}

\noindent
Using the characteristic temperature of the B factor $T^*$, dimensionless thermodynamic parameter $\Sigma^*$, and Nusselt number $Nu$, Eq. \ref{eq_dT_dimension} can be expressed as,

\begin{equation}
	\varDelta T = \frac{\displaystyle T^*}{\displaystyle C_t' Nu
	+ C_s' \frac{\rho_L}{\rho_V} \frac{\Sigma^*}{\sqrt{Re Pr}}},
	\label{eq_dT_dimless}
\end{equation}

\noindent
Equation \ref{eq_dT_dimless} indicates that the temperature depression mainly depends on the characteristic temperature of the B factor $T^*$, and the turbulent convective heat transfer and suppression effect of the evaporation rate determine the magnitude of the temperature depression.
As an expression of the physical mechanism that determines the cavity temperature, the suppression effect equivalent to convective heat transfer $Nu$ is expressed as

\begin{equation}
	S = \frac{\rho_L}{\rho_V} \frac{\Sigma^*}{\sqrt{Re Pr}}.
	\label{eq_Sup}
\end{equation}

\section{Conclusions}
Temperature depression, the temperature difference between the mainstream and cavity, and the indicator of thermodynamic suppression effect resulting from cavitation were modeled and evaluated using the experimental data of temperature measurements of supercavitation in high-temperature water up to 140 \degC.
The following conclusions were drawn:

\begin{enumerate}
	\item Fruman's temperature depression model, which modeled the cooling effect ratio by evaporation and convective heat transfer between the mainstream and cavity, was compared with the experimental data of temperature depression in high-temperature water.
		Then, Fruman's model differed from the experimental data in the high-temperature region of \sidegC{100} and above.
	\item Temperature depression was newly modeled by introducing the suppression effect of evaporative mass flux. 
		By fitting the empirical parameters, the temperature depression was well reproduced in the high-temperature region.
	\item Proposed model was expressed with known parameters; then, it was clear that temperature depression is defined with the characteristic temperature of B factor, Nusselt number, which denote the magnitude of convective heat transfer, and a term representing the suppression effect of evaporative mass flux expressing by dimensionless thermodynamic parameter. 
\end{enumerate}

\nomenclature[A]{\(c_p\)}{specfic heat, \si{J.kg^{-1}.K^{-1}}}
\nomenclature[A]{\(C\)}{coefficient, -}
\nomenclature[A]{\(L\)}{latent heat, \si{J.kg^{-1}}}
\nomenclature[A]{\(L_c\)}{characteristic length, \si{m}}
\nomenclature[A]{\(k\)}{thermal conductivity, \si{W.m^{-1}.K^{-1}}}
\nomenclature[A]{\(h\)}{heat transfer coefficient, \si{W.m^{-2}.K^{-1}}}
\nomenclature[A]{\(T\)}{temperature, \si{K},\si{\degreeCelsius}}
\nomenclature[A]{\(\dot{m}\)}{mass flux, \si{kg.m^{-2}}}
\nomenclature[A]{\(p\)}{pressure, \si{Pa}}
\nomenclature[A]{\(q\)}{heat flux, \si{W.m^{-2}}}
\nomenclature[A]{\(Nu\)}{Nusselt number, -}
\nomenclature[A]{\(Pr\)}{Prandtl number, -}
\nomenclature[A]{\(Re\)}{Reynolds number, -}
\nomenclature[A]{\(U\)}{velocity, \si{m.s^{-1}}}
\nomenclature[B]{\(\Sigma\)}{thermodynamic parameter, \si{m.s^{-1.5}} }
\nomenclature[B]{\(\Sigma^*\)}{dimensionless thermodynamic parameter, - }
\nomenclature[B]{\(\rho\)}{density, \si{kg.m^{-3}} }
\nomenclature[B]{\(\alpha\)}{thermal diffusivity, \si{m^2.s^{-1}} }
\nomenclature[C]{\(b\)}{boundary layer}
\nomenclature[C]{\(sat\)}{saturation}
\nomenclature[C]{\(L\)}{liquid}
\nomenclature[C]{\(V\)}{vapor}
\nomenclature[C]{\(\infty\)}{mainstream}
\nomenclature[C]{\(cav\)}{cavity}
\nomenclature[C]{\(evp\)}{evaporation}

\printnomenclature

\section*{Funding}
This research received no specific grants from any funding agency in the public, commercial, or not-for-profit sectors.

\bibliography{draft}

\end{document}